\documentclass[aps,twocolumn,showpacs,superscriptaddress,groupedaddress]{revtex4}  

\usepackage{amsfonts}
\usepackage[T1]{fontenc}
\usepackage{longtable} 
\usepackage{amsmath}
\usepackage{graphicx} 
\usepackage{dcolumn}  
\usepackage{bm}  
\usepackage{color}  
\usepackage[english]{babel}  
\usepackage{url} 
\usepackage{rotating}
\usepackage[colorlinks=true]{hyperref}
\usepackage{epstopdf}
\usepackage{epsfig}
\usepackage{setspace}
\usepackage{siunitx}
\usepackage{xcolor}

\DeclareGraphicsExtensions{.eps}
\graphicspath{{figures/figures_main/}}

\begin{document} 
\definecolor{ashgray}{rgb}{0.7, 0.75, 0.71}

\title{Search for alternative magnetic tunnel junctions based on all-Heusler stacks}
\author{Worasak Rotjanapittayakul}
\affiliation{Department of Electrical and Computer Engineering, Faculty of Engineering, Thammasat University, Bangkok, 10120, Thailand}
\author{Jariyanee Prasongkit}
\affiliation{Division of Physics, Faculty of Science, Nakhon Phanom University, Nakhon Phanom 48000, Thailand}
\affiliation{Nanotec-KKU Center of Excellence on Advanced Nanomaterials for Energy Production and Storage, Khon Kaen 40002, Thailand.}
\author{Ivan Rungger}
\affiliation{School of Physics, AMBER and CRANN Institute, Trinity College Dublin, Ireland}
\affiliation{Materials Division, National Physical Laboratory, Teddington, TW11 0LW, United Kingdom}
\author{Stefano Sanvito}
\affiliation{School of Physics, AMBER and CRANN Institute, Trinity College Dublin, Ireland}
\author{Wanchai Pijitrojana}
\affiliation{Department of Electrical and Computer Engineering, Faculty of Engineering, Thammasat University, Bangkok, 10120, Thailand}
\author{Thomas Archer}
\email{archert@tcd.ie}
\affiliation{School of Physics, AMBER and CRANN Institute, Trinity College Dublin, Ireland}
\date{\today}

\begin{abstract}
By imposing the constraints of structural compatibility, stability and a large tunneling magneto-resistance, we have identified the 
Fe$_3$Al/BiF$_3$/Fe$_3$Al stack as a possible alternative to the well-established FeCoB/MgO/FeCoB in the search for a novel
materials platform for high-performance magnetic tunnel junctions. Various geometries of the Fe$_3$Al/BiF$_3$/Fe$_3$Al structure 
have been analyzed, demonstrating that a barrier of less than 2~nm yields a tunneling magneto-resistance in excess of 25,000~\% 
at low bias, without the need for the electrodes to be half-metallic. Importantly, the presence of a significant spin gap in Fe$_3$Al
for states with $\Delta_1$ symmetry along the stack direction makes the TMR very resilient to high voltages.  
\end{abstract}


\maketitle


\section{Introduction}
Spin valves displaying large tunnel magneto-resistance (TMR) have undoubtedly revolutionized the electronics industry 
and now form the central component of many technologies, the most successful device being the read heads in hard-disk 
drives~\cite{TMR300}. Importantly, spin valves are set to become the central component of many devices of the future, 
such as magnetic random access memories and spin-torque oscillators. The major breakthrough was the fabrication of 
epitaxial CoFeB/MgO/CoFeB spin valves~\cite{Yuasa,Parkin}, which exploit coherent electronic tunneling~\cite{Butler,Mathon} 
to provide a large TMR even at room temperature. In practical devices the fabrication of CoFeB/MgO/CoFeB spin valves
requires the growth of rather complex thin films stacks, including functional layers (e.g. for magnetic pinning) and seed layers
necessary for the epitaxial growth. It is then desirable to enlarge the available materials platform beyond the CoFeB/MgO system. 
However, despite a large effort in both industry and academia~\cite{spintronics_review}, no junction alternative to 
CoFeB/MgO/CoFeB has been adopted by the community. This is a significant deficiency, since little room is left for tuning the 
materials properties necessary for the development of new applications of the technology. It is therefore imperative to explore 
alternative materials combinations, which offer more freedom to engineer the device properties.

Heusler alloys are a large class of binary ($X_3Z$), ternary ($X_2YZ$) and quaternary ($XX^\prime YZ$) compounds with 
more than 1,500 known members and an impressively wide range of properties~\cite{Heusler}. Many elements can be incorporated 
into the Heusler structure, making it rich and tunable, and as such ideal for developing new technologies. 
 \begingroup
 \squeezetable
 \begin{table}[h!]
 	\caption{\label{table:1}\small{Magnetic tunnel junctions incorporating Heusler alloys electrodes reported to date.
	The TMR is provided for low temperature (LT: 2-16K) and room temperature (RT:$\sim$300K), whenever available.}}
 	\small
 	\begin{ruledtabular}
 		\begin{tabular}{lccc}
 			{MTJ structure}  & TMR$_{\textsc{LT}}$  & TMR$_{\textsc{RT}}$  & Ref.  \\ 
 		\hline
 		  	{Co$_2$MnAlSi/MgO/CoFe}           & 600   & 180    & \cite{Co2MnAlSi}   \\
 			{Co$_2$MnSi/MgO/Co$_2$MnSi}  & 1,995 & 354  &  \cite{Co2MnSi-H}  \\
  			{Co$_2$MnSi/MgO/CoFe}              & 1,049  & 335  & \cite{Co2MnSi-L}  \\
   			{Co$_2$MnGe/MgO/CoFe}            & 376    &  160    & \cite{Co2MnGe-J}  \\
   			{Co$_2$FeAl/Mg$_2$AlO$_4$/CoFe}       & 453   &  280    & \cite{Co2FeAl-H}   \\
     		{Co$_2$FeAl/MgO/Co$_3$Fe}      & -   &  175   & \cite{Co2FeAl-M}   \\
   			{Co$_2$FeAl/MgO/CoFeB}            & -  &  53    &  \cite{Co2FeAl-L}  \\
   			{Co$_2$FeAlSi/MgO/Co$_2$FeAlSi}   & 390  &  220   & \cite{Co2FeAlSi}  \\   			
  			{Co$_2$FeSi/MgO/Co$_3$Fe}      & - &  30        & \cite{Co2FeSi-L}  \\
  			{Co$_2$FeSi/MgO/CoFeB}           & -  &  90   & \cite{Co2FeSi-H}   \\
  			{Co$_2$FeSi/BaO/Fe}                  & -  &  104     & \cite{Co2FeSi-BaO}  \\
  			{Co$_2$CrFeAl/MgO/CoFe}         & 317 &  109     &  \cite{Co2CrFeAl-H}  \\
  			{Co$_2$CrFeAl/MgO/Co$_3$Fe}  & 74 &  42  & \cite{Co2CrFeAl-L}  \\
  			{Co$_2$CrFeAl/MgO/Co$_2$CrFeAl}     & 238 &  60 & \cite{Co2CrFeAl-M}   \\
  			{Mn$_{1.8}$Co$_{1.2}$Ga/MgO/CoFeB}    & -  & 11 &  \cite{Mn2CoGa}  \\
  			{Fe$_2$CrSi/MgO/CoFe}              & - &  8 & \cite{Fe2CrSi}   \\
  			{Fe$_2$CoSi/MgO/Co$_3$Fe}     & 262  &  159 & \cite{Fe2CoSi}  \\
 		\end{tabular}
 	\end{ruledtabular}
 \end{table}
 \endgroup

Several attempts have been made to substitute the magnetic electrodes of the FeCoB/MgO/FeCoB stack with Heusler magnets, 
and successes have been obtained by replacing one or both the electrodes with Co$_2YZ$, where $Y=$~Fe, Mn and $Z=$~Si, 
Al~\cite{H-lead-1,H-lead-2,H-lead-4,Co2Mn1.29Si}.  This body of works is summarized in Table~\ref{table:1}. To the best of our knowledge, 
the highest TMR observed was for Co$_2$MnSi/MgO/Co$_2$MnSi, which displays a TMR ratio of 1,995\% at 4K~\cite{Co2Mn1.29Si}. 
In this system, however, the magnetoresistance is sensitive to temperature with the TMR reducing to 354\% at room temperature~\cite{Co2MnSi-H}. 
Such temperature sensitivity suggests inter-facial magnetic defects or secondary phases, which disrupt the coherent tunneling. Co and Mn 
can directly substitute into the rock-salt MgO structure with formation energies of $-3.0$~eV and $-4.0$~eV, 
respectively~\cite{Aflowlib}, making substitutional Mn in the MgO lattice a likely culprit. 

A second approach has been to construct all Heuslers giant magneto-resistance (GMR) stacks, where the spacer between the 
magnetic electrodes is a metal. The relevant literature is summarized in Table~\ref{table:2}. Although a MR has been demonstrated, 
it was found small for all the known experiments, so that further work is needed to explain these negative results in view of the large 
MR predicted by $ab~initio$ calculations~\cite{all-H-3}. 
 \begingroup
 \squeezetable
 \begin{table}[!ht]
 	\caption{\label{table:2}\small{All-Heusler metallic junctions grown to date. The GMR [\%], $\Delta$RA 
	[m$\Omega\cdot\mu$m$^{2}$] and the method (Exp.=experimental data, {\it ab initio}=theoretical prediction) 
	are given.}}
 	\small
 	\begin{ruledtabular}
 		\begin{tabular}{lcccc}
 			{all Heusler structure}  & GMR & $\Delta$RA & method  & Ref.  \\ 
 			\hline
 			{Co$_2$MnGe/Rh$_2$CuSn/Co$_2$MnGe}   & 6.7 & 4  &Exp.&\cite{all-H-1}   \\
 			{Co$_2$MnSi/Ni$_2$NiSi/Co$_2$MnSi }  & -  &  0.24 &Exp.& \cite{all-H-2}  \\
 			{Co$_2$CrSi/Cu$_2$CrAl/Co$_2$CrSi }  & $\sim10^{6}$ &  - &$ab~initio$&\cite{all-H-3}  \\
 		\end{tabular}
 	\end{ruledtabular}
 \end{table}
 \endgroup

The question that we answer to in this paper is the following: given the wide variety of properties available in the Heusler class, is it 
possible to create an all-Heusler TMR junction with materials alternative to the Fe/MgO system? In this work we will use simple 
design concepts and \textit{ab initio} calculations to screen candidates based on the symmetry filtering mechanism, which has been 
so successful for the Fe/MgO junction. Our analysis returns us the Fe$_3$Al/BiF$_3$ system as a promising stack for large
magnetoresistance with a strong TMR retention at high bias. The paper is organized as follows. We open our discussion by explaining 
the criteria that have brought us to focus on a particular Heusler alloys stack, by looking first at the barrier and then at the magnetic
electrodes. Then we move to discuss the transport properties of several Fe$_3$Al/BiF$_3$/Fe$_3$Al junctions with different barrier
thicknesses. We first look at the zero-bias properties and then move to the finite-bias ones. Finally we conclude.

\section{Screening the materials}
\subsection{The tunnel barrier}
In order to propose a new junction we must satisfy a number of constraints, which we will use to screen candidate materials 
combinations. Firstly, the barrier material must be a robust insulator and therefore must have a large band gap, $E_\mathrm{g}$. 
If we use a cut off band gap of 2.5~eV, we will reduce the number of the candidate Heusler materials from over 300,000, 
(these include those reported in literature and the hypothetical ones contained in the AFLOW.org database~\cite{Aflow.org}), 
to just 26. Notably, only 4 of these have been grown experimentally before, the remaining 22 are only predicted from \textit{ab initio} 
calculations~\cite{Aflow.org}. The 26 barrier candidates are shown in Table~\ref{table:3}. Note that the band gaps reported here 
are computed by density functional theory (DFT) in the generalized gradient approximation (GGA), therefore they are expected to 
be significantly smaller than the true quasi-particle band gap. As such our $E_\mathrm{g}\geq$2.5~eV criterion effectively selects 
insulators with a band gap, which is likely to be significantly larger than 2.5~eV.
\begingroup
\squeezetable
\begin{table}[!ht]
\caption{\label{table:3}\small{All possible insulating Heusler materials having a wide band-gap, $E_\mathrm{g}\geq$2.5~eV.
The Strukturbericht (SB) symbols, lattice constant ($a_0$[\AA]), band gap ($E_\mathrm{g}$[eV]), tetragonal ratio ($c/a_0$) 
and the method with which they have been investigated, are given (Exp.=experimental data, {\it ab initio}=theoretical prediction 
from AFLOW.org).}}
\small
\begin{ruledtabular}
\begin{tabular}{lcccccc}
{Material}  & SB & $a_0$  & $E_\mathrm{g}$ & $c/a_0$ &method & Ref. \\ \hline
BaBrCl     & C1$_b$    & 7.604   & 3.476  &-& $ab~initio$ &  \cite{Aflow.org} \\
{BiF$_3$}   & D0$_3$   & 5.861  & 5.100  &1.0& Exp. & \cite{lat-BiF3, exp_gap_BiF3,mm} \\

BrClSr    & C1$_b$    & 7.262  & 4.670  &-& $ab~initio$ & \cite{Aflow.org} \\
BrClPb    & C1$_b$    & 7.251    & 3.090  &-& $ab~initio$ & \cite{Aflow.org} \\
BrCaCl    & C1$_b$    & 6.973   & 4.386  &-& $ab~initio$ & \cite{Aflow.org} \\
BrHgK   & C1$_b$    & 7.948   & 3.253  &-& $ab~initio$ & \cite{Aflow.org} \\
Br$_2$KLi   & L2$_1$   & 7.647   & 3.313   &-& $ab~initio$ & \cite{Aflow.org} \\
Br$_2$KNa   & L2$_1$    & 7.784    & 3.337   &-& $ab~initio$ & \cite{Aflow.org} \\
Br$_2$KTl   & L2$_1$     & 8.083    & 3.330  &-& $ab~initio$ & \cite{Aflow.org} \\
Br$_2$LiNa   & L2$_1$    & 7.251   & 3.045  &-& $ab~initio$ & \cite{Aflow.org} \\

Cl$_2$GaK     & L2$_1$    & 7.493    & 3.424  &-& $ab~initio$ & \cite{Aflow.org} \\
Cl$_2$GaNa   & L2$_1$    & 7.198   & 3.032  &-& $ab~initio$ & \cite{Aflow.org} \\
Cl$_2$InK   & L2$_1$    & 7.718    & 3.154  &1.2& $ab~initio$ & \cite{Aflow.org,mm} \\
Cl$_2$KLi   & L2$_1$    & 7.230    & 4.293  &1.0& $ab~initio$ & \cite{Aflow.org,mm} \\
Cl$_2$KNa   & L2$_1$    & 7.367    & 4.277   &-& $ab~initio$ & \cite{Aflow.org} \\
Cl$_2$KTl   & L2$_1$   & 7.749   & 3.801  &-& $ab~initio$ & \cite{Aflow.org} \\
Cl$_2$LiNa   & L2$_1$  & 6.793    & 4.194  &-& $ab~initio$ & \cite{Aflow.org} \\
Cl$_2$LiTl   & L2$_1$   & 7.397   & 3.281  &1.2& $ab~initio$ & \cite{Aflow.org,mm} \\

ClHgK     & C1$_b$  & 7.771  & 3.531 &-& $ab~initio$ & \cite{Aflow.org} \\
ClKZn    & C1$_b$   & 7.637   & 3.107  &-& $ab~initio$ & \cite{Aflow.org} \\
ClHgK   & C1$_b$    & 7.778  & 3.143 &-& $ab~initio$ & \cite{Aflow.org} \\

{LiMgN}     & C1$_b$       & 4.955     & 3.200 &1.0& Exp. & \cite{LiMgN} \\
{LiMgP}     & C1$_b$     & 6.005       & 2.430 &1.0& Exp. &  \cite{LiMgP} \\
LiNaS  & C1$_b$   & 6.100   & 3.120  &1.0& $ab~initio$  & \cite{LiNaZ}\\
LiNaSe  & C1$_b$   & 6.390 & 2.700 &1.0& $ab~initio$ & \cite{LiNaZ} \\

{TaIrGe}    & C1$_b$  & 5.967  & 3.360 &1.0&  Exp.  & \cite{exp_TaIrGe} \\

\end{tabular}
\end{ruledtabular}
\end{table}
\endgroup

Next we consider the transport properties of the tunnel barrier. In epitaxial spin valves the magnitude of the TMR is determined 
by the symmetry matching between the evanescent wave-functions in the insulating barrier and the Bloch wave-functions for 
majority and minority spins in the magnetic electrodes. In particular the TMR will be large if such wave-function symmetry match 
occurs for only one of the two spin species, i.e. if only one of the two spin species is transmitted with large 
probability~\cite{Butler,Mathon}.

In order to further screen these candidate barrier materials we have performed electronic structure calculations using 
self-interaction-corrected~\cite{Asic-Method1,Asic-Method2} DFT as implemented in the atomic-orbital-based code 
\textsc{Siesta}~\cite{SIESTA}. In general the inclusion of self-interaction corrections drastically improve the band gap 
of a broad range of insulators, and here it brings the calculated gap of the chosen Heusler barriers within 10\% from the 
experimental one. Core electrons are treated with norm-conserving relativistic Troullier-Martin pseudopotentials~\cite{PSP}, 
while multi-$\zeta$ numerical atomic orbitals are used to represent the electron density and all the operators. Total energies 
are computed on a uniform real-space grid with an equivalent cutoff of 600~Ry, while the primitive unit cells are sampled 
with a 8$\times$8$\times$8 $k$-point mesh over the Brillouin zone. The linear response conductance is calculated with 
the DFT-based non-equilibrium Green's functions code \textsc{Smeagol}~\cite{smeagol2,SigmaMethod}, where the typical 
$k$-point sampling for a given heterojunction is 100$\times$100$\times$1.
\begingroup
\squeezetable
\begin{table}[!ht]
\caption{\label{table:x}\small{Insulating Heusler materials with a band gap $E_\mathrm{g}\geq$2.5 eV. The Strukturbericht (SB) symbols, 
lattice constant ($a_{\texttt{exp}}$[\AA]) and experimental band gap ($E_\mathrm{g}$[eV]) are given. In the final column 
we report the symmetry of the slowest decaying state along the [001] direction, as calculated in this work.}}
\small
\begin{ruledtabular}
\begin{tabular}{lcccc}
    Material  & SB & $a_{\mathrm{exp}}$  & $E_\mathrm{g}$ & Symmetry  \\ \hline
    {BiF$_3$} & D0$_3$ \cite{lat-BiF3}   & 5.861 \cite{lat-BiF3}   & 5.10 \cite{exp_gap_BiF3} & $\Delta^{\mathrm{CB-VB}}_1$ \\
    {LiMgN}   & C1$_b$ \cite{LiMgN}    & 4.955 \cite{LiMgN}    & 3.20 \cite{LiMgN}      & $\Delta^{\mathrm{CB}}_{5}$, $\Delta^{\mathrm{VB}}_1$ \\
    {LiMgP}   & C1$_b$ \cite{LiMgP}    & 6.005 \cite{LiMgP}    & 2.43 \cite{LiMgP}      & $\Delta^{\mathrm{CB}}_{5}$, $\Delta^{\mathrm{VB}}_1$ \\ 
    {TaIrGe}  & C1$_b$ \cite{exp_TaIrGe} & 5.967 \cite{exp_TaIrGe} & 3.36 \cite{exp_TaIrGe}   & $\Delta^{\mathrm{CB}}_1$, $\Delta^{\mathrm{VB}}_2$
\end{tabular}
\end{ruledtabular}
\end{table}
\endgroup

For each of the insulators we determine the symmetry of the slowest decaying state along the [001] direction, and we restrict 
ourselves to the experimentally verified insulating Heusler alloys, namely BiF$_3$, LiMgN, LiMgP and TaIrGe. Our results are 
presented in Table~\ref{table:x}, where we list the experimental structural parameters and quasi-particle band gap, together with
the symmetry of the evanescent wave function with the slowest decay across the barrier. Notably, while for BiF$_3$ there is only
one low-lining complex band crossing the band gap, this is not the case for the other three alloys. In fact, for LiMgN, LiMgP and 
TaIrGe the symmetry of the valence band maximum (VBM) and conduction band minimum (CBM) are different. This means that 
there is not a single complex band bridging the band gap, since the one starting at the VBM (CBM) does not end at the CBM (VBM).
As such, the symmetry of the slowest decaying state across the barrier depends upon the exact position of the Fermi level in the
hypothetical junction, namely on the band alignment. This situation is not desirable in a tunneling junction~\cite{Nuala}. When all these 
features are brought together, BiF$_3$ appears as our best candidate. Its band structure is illustrated in Fig.~\ref{fig:banddd}\textcolor{red}{(d)}. 

BiF$_3$ is the naturally occurring mineral {\it Gananite}, which has been reported to have a D0$_3$ structure 
and a lattice parameter of 5.861\AA~\cite{lat-BiF3}. The F atoms occupy the 4$a$, 4$b$ and 4$c$ Wyckoff 
positions, while Bi is accommodated in the 4$d$. Gananite is a wide band-gap insulator with an experimentally 
observed optical gap of $\sim$5.10 eV~\cite{exp_gap_BiF3}. Theoretical band gaps of 3.81~eV and 3.94~eV were 
calculated with the LDA (at the LDA lattice constant of $a^{\mathrm{\tiny{LDA}}}_0$=5.865\AA)~\cite{gapBiF3-FP-LMTO}
and the GGA (at the GGA lattice constant of $a^{\mathrm{\tiny{GGA}}}_0$=5.860\AA)~\cite{GapBiF3,gapBiF3_1}, respectively.
In this work the atomic self-interaction correction (ASIC) scheme built on top of the LDA returns us a value of 5.25~eV 
($a^{\mathrm{\tiny{ASIC}}}_0$=5.836\AA).

\subsection{The magnetic electrodes}
We now move to select the magnetic materials to be used as electrodes. A crude screening criterion is that the magnetic 
electrodes must be made of materials having a magnetic ordering temperature significantly higher than room temperature. 
Here we have chosen the cutoff to be 700~K, a value that should be sufficient to ensure little magnetization degradation for
temperatures around room temperature. Such cutoff temperature reduces the number of candidates to the 20 alloys listed in 
Table~\ref{table:4}.
\begingroup
\squeezetable
\begin{table}[htp!] 
	\caption{\label{table:4}\small{Magnetic Heusler materials with a $T_\mathrm{C}$ greater than 700~K considered as potential 
	electrode. We report the Strukturbericht (SB) symbols, the experimental  lattice constant ($a_{\mathrm{exp}}$[\AA]),  
	the Curie temperature ($T_\mathrm{C}$ [K]), and the magnetic order, FM=ferromagnetic, HFM=half-metal. Here ``ferri''
	means that the magnetic order is ferrimagnetic, although the electronic structure is that of a half-metal.}}
	\small
	\begin{ruledtabular}
		\begin{tabular}{ lccccc }
			Material & SB & $a_{\mathrm{exp}}$  & $T_\mathrm{C}$ & Magnetic ground state & Ref.\\ 
			\hline 
			{Fe$_3$Al}     &  D0$_3$  &  5.793  &  713 & FM &~\cite{exp_Fe3_Al-Si} \\
			{Fe$_3$Si}     &  D0$_3$  &  5.553  &  840 & FM  & ~\cite{exp_Fe3_Al-Si} \\
			{Fe$_2$CoGe}  &  D0$_3$  &  5.780  &  925 & FM &  ~\cite{m_exp_Fe2CoGe} \\ 
			{Fe$_2$CoSi}  &  D0$_3$  &  5.645  &  1,025 & FM &  ~\cite{m_exp_Fe2CoSi} \\   
			{Fe$_2$CuAl}  &  A$_2$   &  5.830  &  875 & FM & ~\cite{m_exp_Fe2CoGe}  \\
			{Fe$_2$NiGe}  &  A$_2$   &  5.761  &  750 & FM & ~\cite{m_exp_Fe2CoGe}  \\
			{Fe$_2$NiAl}  &  L2$_1$  &  5.778  &  965 & FM & ~\cite{m_exp_Fe2NiAl}   \\
			{Fe$_2$NiSi}  &  D0$_3$  &  5.671  &  755 & FM & ~\cite{m_exp_Fe2NiAl}  \\
			{Co$_2$MnAl}  &  B2  &  5.671  &  710 & HMF  & ~\cite{exp_Co2MnAl} \\
			
			{Co$_2$MnSi}   &  L2$_1$  &  5.655  &  985 & HMF & ~\cite{exp_Co2MnSi} \\
			{Co$_2$MnGe}  &  L2$_1$  &  5.749  &  905 & HMF & ~\cite{Co2MnGe}   \\
			{Co$_2$MnSn}  &  L2$_1$  &  6.000  &  829 & HMF & ~\cite{Co2MnSn-1, Co2MnSn-2}  \\
			{Co$_2$FeSi}   &  L2$_1$  &  5.640  &  1,100 & HMF & ~\cite{Co2FeX-1, Co2FeSi-1, Co2FeSi-2}  \\
			{Co$_2$FeAl}  &  B2      &  5.737  &  1,000 & HMF & ~\cite{Co2FeX-1, Co2FeAl-1, Co2FeAl-2}  \\
			{Co$_2$FeGa}  &  L2$_1$  &  5.751  &  1,100 & HMF & ~\cite{Co2FeX-1, Co2MnSn-1}  \\
			{Co$_2$FeGe}  &  L2$_1$  &  5.743  & 981 & HMF & ~\cite{Co2FeX-1, Co2FeGe}   \\
			{Co$_2$CrSi}  &  L2$_1$  &  5.647  &  747 & HMF & ~\cite{Co2CrSi}  \\
			{NiMnSb}  &  C1$_b$  &  5.903  &  730  & HMF & ~\cite{NiMnSb}   \\			
			{Mn$_2$VAl}  &  L2$_1$  &  5.920 &  760  & HMF (ferri)  &  ~\cite{Mn2VAl} \\
			{Mn$_2$CoGa}  &  L2$_1$  &  5.873 &  740  & HMF (ferri)  &  ~\cite{Mn2CoGa-X} \\
		\end{tabular}
	\end{ruledtabular}
\end{table}
\endgroup

Secondly, there should be a good lattice match between the magnetic electrodes and the insulator. This is a necessary
condition to ensure the epitaxial grow of the stack, which in turn is necessary for the spin filtering. We set the tolerance
for the lattice match to less than 1.5\%. Such match can be achieved either by having a one-to-one match between the 
insulator and the magnet (the two share the same crystallographic axes), or by rotating one of them by \ang{45} in the 
plane of the stack (here we consider only the (100) growth direction). Table~\ref{table:miss} presents all the possible
electrode/barrier combinations having a lattice mismatch smaller than 1.5\%, with the \ang{45}-rotated epitaxial 
structures being in grey colour. 
\begingroup
\squeezetable
\begin{table}[htp!] 
\caption{\label{table:miss}\small{Materials combinations presenting a lattice mismatch smaller than 1.5\%. This can be
obtained with the barrier and the magnet sharing the same crystallographic axes, or by rotating one of them by
\ang{45}(coloured gray). }}
\small
\begin{ruledtabular}
		\begin{tabular}{ l|ccccccccccccccccccccccccc }
&\rotatebox{90}{ BiF$_3$  }&\rotatebox{90}{ TaIrGe  }&\rotatebox{90}{ LiMgP  }&\rotatebox{90}{ Cl$_2$KTl  }&\rotatebox{90}{  Br$_2$KLi }&\rotatebox{90}{ ClHgK }&\rotatebox{90}{ Br$_2$KNa  }&\rotatebox{90}{ BrHgK  }&\rotatebox{90}{ Br$_2$KTl} \\
\hline
Fe$_3$Si&&&&\color{ashgray}1.3&\color{ashgray}1.1&\color{ashgray}1.0&\color{ashgray}0.9&\color{ashgray}1.2& \\
Co$_2$FeSi&&&&&&&&\color{ashgray}0.4&\color{ashgray}1.3 \\
Fe$_2$CoSi&&&&&&&&\color{ashgray}0.4&\color{ashgray}1.2 \\
Co$_2$CrSi&&&&&&&&\color{ashgray}0.5&\color{ashgray}1.2 \\
Co$_2$MnSi&&&&&&&&\color{ashgray}0.6&\color{ashgray}1.1 \\
Fe$_2$NiSi&&&&&&&&\color{ashgray}0.9&\color{ashgray}0.8 \\
Co$_2$MnAl&&&&&&&&\color{ashgray}0.9&\color{ashgray}0.8 \\
Co$_2$FeAl&&&&&&&&&\color{ashgray}0.4 \\
Co$_2$FeGe&&&&&&&&&\color{ashgray}0.5 \\
Co$_2$MnGe&&&&&&&&&\color{ashgray}0.6 \\
Co$_2$FeGa&&&&&&&&&\color{ashgray}0.6 \\
Fe$_2$NiGe&1.4&&&&&&&&\color{ashgray}0.8 \\
Fe$_2$NiAl&1.4&&&&&&&&\color{ashgray}1.1 \\
Fe$_2$CoGe&1.4&&&&&&&&\color{ashgray}1.1 \\
Fe$_3$Al&1.2&&&&&&&&\color{ashgray}1.4 \\
Fe$_2$CuAl&0.5&&&&&&&& \\
Mn$_2$CoGa&0.2&&&&&&&& \\
NiMnSb&0.7&1.1&&&&&&& \\
Mn$_2$VAl&1.0&0.8&1.4&&&&&& \\
Co$_2$MnSn&&0.6&0.1&&&&&& \\
		\end{tabular}
	\end{ruledtabular}
\end{table}
\endgroup

From the table it is easy to note that there are only eight magnets presenting a lattice mismatch smaller than 1.5\% with our
chosen insulator, BiF$_3$. Two of these, Mn$_2$CoGa and Mn$_2$VAl, are Mn$_2$-based Heusler alloys, which we exclude 
from further analysis. The reason for such exclusion is that often the ground state of Mn$_2$-type alloys presents a complex magnetic 
structure with ferrimagnetic order between the crystallographic inequivalent Mn ions (e.g. see Mn$_3$Ga~\cite{Karsten}). This 
is a situation, which is not suitable for a spin-valve. We also exclude the half-Heusler, NiMnSb, which has a half-metallic
electronic structure, but it is prone to disorder that strongly modifies its magnetic properties~\cite{doi:10.1063/1.372550}.
The electronic structure of the remaining five electrode compounds has been calculated, the symmetry of the states at the 
Fermi-level has been analysed and it is summarized in table~\ref{table:y}. Given the symmetry of the relevant complex band
in BiF$_3$, the electrodes must present bands with $\Delta_1$ symmetry at the Fermi level for only one spin channel along 
the [001] direction. All the five remaining candidates meet this criteria.

\begingroup
\squeezetable
\begin{table}[htp!] 
\caption{\label{table:y}\small{Magnetic Heuslers considered as potential electrodes. The Strukturbericht (SB) symbols, 
the experimental  lattice constant ($a_{\mathrm{exp}}$[\AA]), the magnetic moment per formula unit 
($\mu_\mathrm{S}$[$\mu_{\mathrm{B}}$/f.u.]) and the Curie temperature ($T_\mathrm{C}$ [K]) are given. In two 
final columns we show the band symmetry ($\Delta^{\sigma}_{[001]}$ for spin $\sigma=\uparrow, \downarrow$) across 
the Fermi level along the [001] direction, which have been calculated in this work. For Fe$_3$Al no minority spin band 
crosses the Fermi level along the [001] direction (note that globally Fe$_3$Al is not a half-metal, but it is along [001]).}}

\small
\begin{ruledtabular}
\begin{tabular}{ l|c|c|c|c|c|c }
  Material & SB & $a_{\mathrm{exp}}$  & $\mu_\mathrm{S}$ & $T_\mathrm{C}$ & {$\Delta^{\uparrow}_{[001]}$} & $\Delta^{\downarrow}_{[001]}$ \\ 
  \hline 
  {Fe$_3$Al} ~\cite{exp_Fe3_Al-Si}    &  D0$_3$  &  5.793 &  5.10  &  713    & $\Delta_1$,$\Delta_5$ & - \\
  {Fe$_2$CoGe} ~\cite{m_exp_Fe2CoGe}  &    D0$_3$     &  5.780 &  5.40  &  925    & $\Delta_1$,$\Delta_5$ & $\Delta_5$ \\
  {Fe$_2$CuAl} ~\cite{m_exp_Fe2CoGe}  &    A$_2$     &  5.830 &  3.30  &  875    & $\Delta_1$,$\Delta_5$ & $\Delta_{5}$ \\
  {Fe$_2$NiGe} ~\cite{m_exp_Fe2CoGe}  &    A$_2$     &  5.761 &  4.29  &  750    & $\Delta_1$,$\Delta_5$ & $\Delta_5$,$\Delta_2$,$\Delta_{2'}$ \\   
  {Fe$_2$NiAl} ~\cite{m_exp_Fe2NiAl}  &    L2$_1$     &  5.778 &  4.46  &  965    & $\Delta_1$,$\Delta_5$ & $\Delta_{2}$,$\Delta_{2'}$ \\ 
\end{tabular}
\end{ruledtabular}
\end{table}
\endgroup
%
%
When looking at the electronic structure of the five remaining Fe-containing magnetic Heusler alloys we notice that all
of them present bands at the Fermi level with both $\Delta_1$ and $\Delta_5$ symmetry in the majority ($\uparrow$) 
channel, while the symmetry of the minority one ($\downarrow$) differentiates them. In two cases, Fe$_2$CoGe and 
Fe$_2$CuAl, the minority Fermi surface is dominated by the $\Delta_5$ symmetry, while for other two, Fe$_2$NiAl and
Fe$_2$NiGe, both $\Delta_2$ and $\Delta_{2^\prime}$ bands are present (in the case of Fe$_2$NiGe there is also a
$\Delta_5$ one). Fe$_3$Al sets a case on its own, since there is a spin-gap in the minority band. Note that this is not
a complete spin gap, namely Fe$_3$Al is not a half-metal, but it is present along the (100) direction. For this reason, 
among the different possibilities, we have then chosen Fe$_3$Al as electrode material. Fe$_3$Al has high $T_\mathrm{C}$ 
(713~K)~\cite{exp_Fe3_Al-Si}, and only a 1.2\% lattice mismatch to BiF$_3$. It has a D0$_3$-structure ($Fm\bar{3}m$), 
Fe(I) atoms occupy the Wyckoff positions 4$a$ ($0,0,0$) and 4$b$ ($\frac{1}{2},\frac{1}{2},\frac{1}{2}$), while Fe(II) and 
Al atoms occupy the 4$c$ ($\frac{1}{4},\frac{1}{4},\frac{1}{4}$) and 4$d$ ($\frac{3}{4},\frac{3}{4},\frac{3}{4}$) ones, respectively. 
\begin{figure}[ht!]
\begin{centering}
	\includegraphics[width=1.0\linewidth]{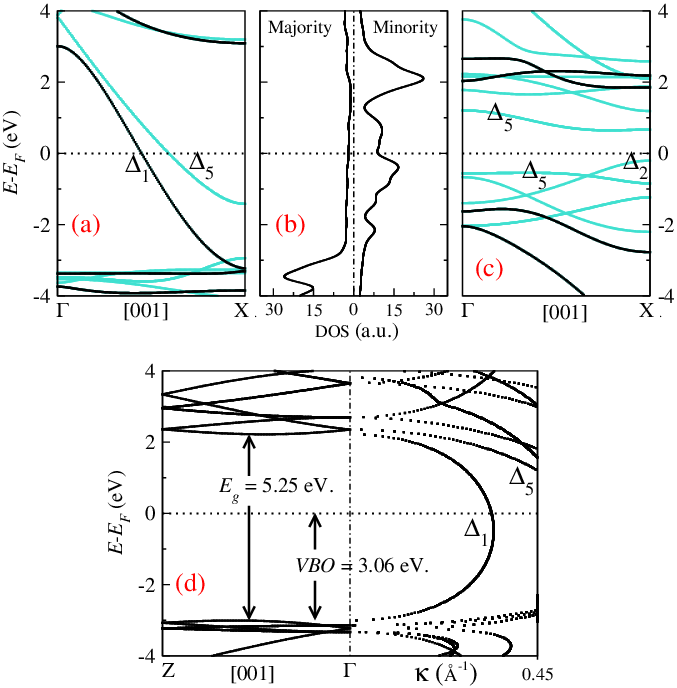}
	\caption{\label{fig:banddd}\small{Electronic structure of Fe$_3$Al and BiF$_3$ along the device stack direction, 
	[001]. Panels (a), (b) and (c) are the majority band structure, the density of states and the minority band structure 
	for D0$_3$-Fe$_3$Al, respectively. The bold lines represent the $\Delta_1$ bands. Panel (d) displays the complex 
	band structure of the bulk BiF$_3$.
	}}
\end{centering}
\end{figure}

In Figs.~\ref{fig:banddd}(a, c) we present the band structure for majority and minority spins along [001] (the proposed 
stack orientation). As we have seen, BiF$_3$ filters states with $\Delta_1$ symmetry, which are present in Fe$_3$Al 
only for the majority band. In fact, along the [001] direction ($\Gamma\rightarrow$X in $k$-space) at $E_\mathrm{F}$ 
there is a wide $\Delta_1$ band originating from the Al 3$s$ and Fe 4$s$ states in the majority spin channel [Fig.~\ref{fig:banddd}(a)], 
and a band gap in the minority one [Fig.~\ref{fig:banddd}(c)]. The first $\Delta_1$ contributions for the minority spin 
appear at $\pm$1.5~eV from $E_\mathrm{F}$, providing a ~3~Volt window in which the device is expected to show a large 
TMR. Note that, as already mentioned, the material is not half-metal as the gap in the minority channel is only along the specific 
$\Gamma\rightarrow$X direction, as shown in Fig.~\ref{fig:banddd}(b).

\section{The all-Heusler F$\mathbf{e}_3$A$\mathbf{l}$/B$\mathbf{i}$F$_3$/F$\mathbf{e}_3$A$\mathbf{l}$ spin valve}
\subsection{Zero bias properties}
\begin{figure}[!ht]
\centering
	\includegraphics[width=1.0\linewidth]{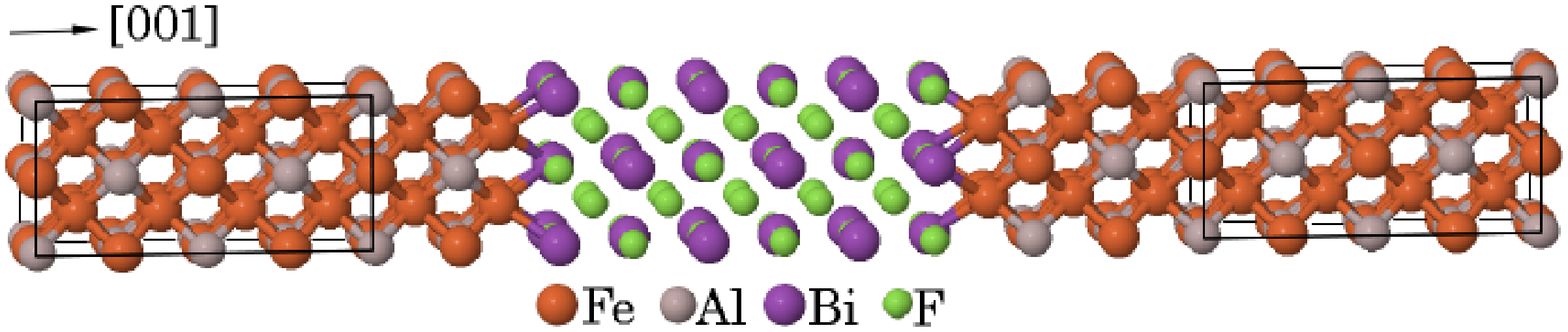}
	\caption{\label{Schemetic}\small{Atomic structure of the all-Heusler Fe$_3$Al/BiF$_3$/Fe$_3$Al spin valve. 
	The system is periodic in the plane orthogonal to [001], which defines the transport direction.}}
\end{figure}

%
The all-Heusler Fe$_3$Al/BiF$_3$/Fe$_3$Al spin valve is constructed by stacking Fe$_3$Al(001) on BiF$_3$(001), 
as shown in Fig.~\ref{Schemetic}. Its transport properties are now systematically investigated.

For our transport calculations the in-plane lattice constant is fixed at $a_{0}$=5.836~\AA, equivalent to the theoretical cubic 
lattice constant of bulk BiF$_3$. Fixing the in-plane lattice constant induces a small tetragonal distortion in the semi-infinite 
Fe$_3$Al(001) leads with $c/a_{0}$=1.124 (the cell is relaxed to a forces tolerance of 10~meV/\AA). Such a distortion has 
negligible effects on the electronic structure of the electrodes. The interface energy, corrected for basis set superposition error, 
is found to be 3.78~J/m$^{2}$. To put this in context, the computed Fe/MgO interface energy is reported to be 2.52~J/m$^{2}$~\cite{SurfE}, 
namely the Fe$_3$Al/BiF$_3$ interface seems to be stronger than the Fe/MgO one. The Fermi level of the 
junction is found to lie approximately in the middle of the BiF$_3$ band gap, with a valence band offset of 3.06~eV, as shown in 
Fig.~\ref{fig:banddd}(d).
  
\begin{figure}[h]
\begin{centering}
	\includegraphics[width=1\linewidth]{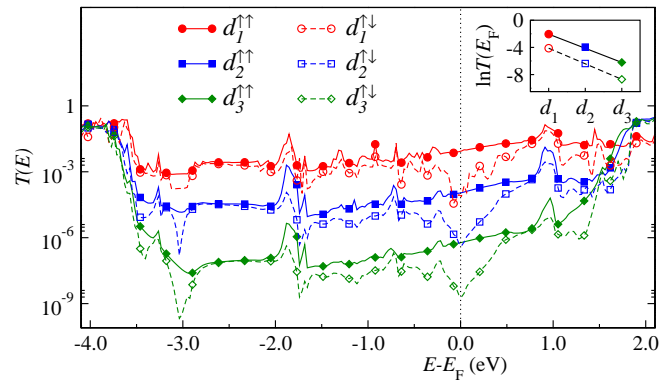}
	\caption{\label{fig:trc_bias}\small{Zero-bias transmission coefficient as a function of energy for the two configurations, where the
	magnetization vectors of the electrodes are either parallel ($\uparrow\uparrow$) or anti-parallel ($\uparrow\downarrow$) to each
	other. Results are presented for three barrier thicknesses: $d_1$=13.10~\AA\ (red lines), $d_2$=18.94~\AA\ (blue lines) and
	$d_3$=24.77~\AA\ (green lines). The inset shows the barrier thickness dependence of $T(E_\mathrm{F})$. All results are obtained 
	with a 100$\times$100 $k$-mesh.}}
	\end{centering}
\end{figure}

Electronic transport is calculated for three junctions with different BiF$_3$ thicknesses, respectively of 13.10~\AA, 18.94~\AA\ 
and 24.77~\AA. The zero-bias transmission coefficients as a function of energy, $T(E)$, are shown in Fig.~\ref{fig:trc_bias} and 
clearly demonstrate that there is an exponential reduction of the transmission with the barrier thickness, confirming that the 
transport mechanism is indeed tunneling with little contribution from possible interface states. For the thinnest junction, 13.10~\AA, 
there is a some structure in $T(E)$ for the anti-parallel configuration around $E_\mathrm{F}$. This is the result of the 
$\Delta^{\downarrow}_{5,2,2'}$ bands not being fully filtered, but it disappears for thicker junctions, for which $T(E)$ around
$E_\mathrm{F}$ is smooth.

\begin{figure}[h]
\begin{centering}
	\includegraphics[width=1\linewidth]{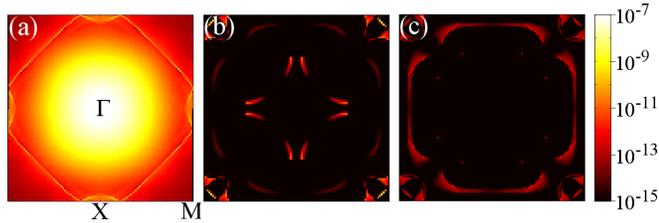}
	\caption{\label{fig:TRC_new}\small{$k_{\parallel}$-resolved transmission coefficient at the Fermi level for the all-Heusler junction 
	with a BiF$_3$ barrier of 18.94~\AA.  (a) majority spins parallel configuration, (b) minority spins parallel configuration, (c) anti-parallel
	configuration. All results are obtained for a 100$\times$100 $k$-mesh.}}
	\end{centering}
\end{figure}

The various transmission coefficients for the [001] direction calculated at the Fermi level are plotted in Fig.~\ref{fig:TRC_new} as a 
function of the $k$-vector in the two-dimensional Brillouin zone orthogonal to the transport direction. For the parallel configuration 
the transmission is dominated by the majority spins and a $k$-region around the $\Gamma$-point, while for the minority band
and for the anti-parallel configuration the transmission is small and originates from narrow pockets of $k$-vectors away from $\Gamma$. 
This further confirms that the transport is dominated by the $\Delta_1$ symmetry, present only for the majority spins. Importantly, 
the relative contribution to the total current of the majority spin channel relatively to the minority one in the parallel configuration will 
exponentially grow as the barrier thickness increases, meaning that for barriers thick enough the Fe$_3$Al/BiF$_3$ system behaves
as a half-metal, exactly as Fe/MgO.


\subsection{Finite-bias properties}
\begin{figure}[!ht]
   \begin{centering}
	\includegraphics[width=1\linewidth]{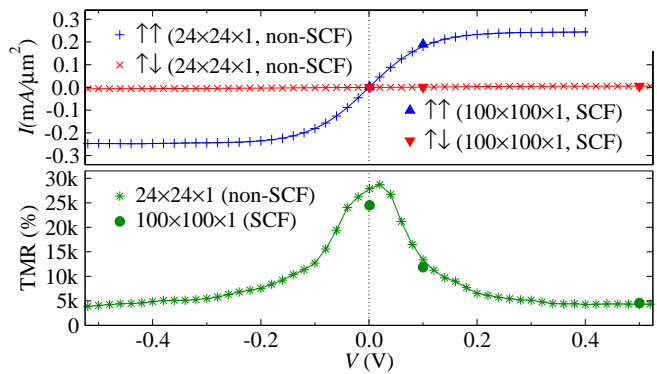}
	\caption{\label{fig:TMR}\small{Total current density, $I$, and TMR versus bias voltages, $V$, for the parallel and anti-parallel magnetic 
	configuration of the junction. Results are presented for a non-self-consistent calculation (the potential drop is not calculated self-consistently) 
	using a 24$\times$24 $k$-grid and confirmed by a self-consistent one obtained with a 100$\times$100 mesh (closed symbols at voltages
	$V=0, 0.1, 0.5$~V). The BiF$_3$ thickness is 18.94~\AA.}}
	\end{centering}
\end{figure}

For the 18.94~\AA-thick junction we have calculated the current and the TMR as a function of bias [see Fig.~\ref{fig:TMR}]. 
Calculations are performed on a 24$\times$24$\times$1 $k$-point mesh non-self consistently (the potential drop is not self-consistently
evaluated, see~\cite{Nutta}), and have been verified against a 100$\times$100$\times$1 mesh for a self-consistent calculation at 
a few selected biases (0, 0.1 and 0.5~Volt). We find that the $I$-$V$ characteristic of the parallel configuration is approximately linear 
at low bias and then saturates at about 0.2~Volt to a value of 0.25~mA/$\mu$m$^2$. Since the same curve for the anti-parallel 
configuration is flat and the current is small the TMR as a function of bias decays from the $V\sim0$ value of 25,000~\% to about 
5,000\% at $|V|>$~0.2~Volt [see below for $T(E;V)$]. This is indeed a very encouraging result since an extremely large TMR can be 
reached for a 2~nm think barrier, and larger values can be obtained by making the barrier thicker. We must note that the theoretical 
TMR is for the perfect junction and demonstrates that symmetry filtering is the dominant mechanism. The actual TMR of any junction 
will depend on secondary phases or defects at the interface so the values observed here should be considered an upper limit. 
\begin{figure}[!ht]
\begin{centering}
\includegraphics[width=1\linewidth]{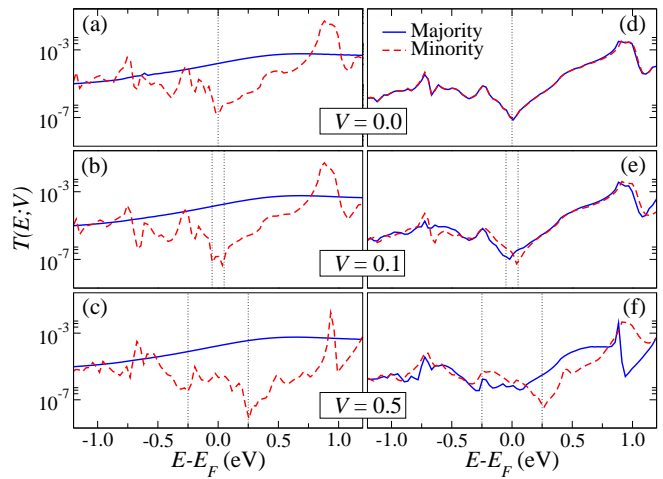}
\caption{\label{fig:Bias}\small{Self-consistently calculated finite-bias transmission coefficient $T$($E$;$V$) as a function of 
energy for the (a-c) parallel and (d-f) anti-parallel magnetic configuration of the electrodes. The vertical dotted-lines are located 
at $E=E_\mathrm{F}\pm eV/2$, namely they enclose the bias window. Note that the transition remains spin degenerate only in 
the case of zero-bias for the anti-parallel configuration.}}
\end{centering}
\end{figure}

The transmission of the Fe$_3$Al/BiF$_3$/Fe$_3$Al junction has been self-consistently calculated at 0~Volt, 0.1~Volt and 0.5~Volt. 
In figure~\ref{fig:Bias} we present the transmission coefficient for each bias step, $T(E;V)$. The behavior of the junction can be 
understood by considering the $\Delta_1$ filtering of BiF$_3$ and the band structure of the Fe$_3$Al electrodes. We see that
when the magnetizations of the electrodes are parallel, $T(E;V)$ for the majority spins is a smooth function of the energy, since 
the transmission originates from $\Delta_1$ band. At the same time there is no minority spin bands at the Fermi-level along the [001] 
direction, resulting in a strongly suppressed minority transition around $E_\mathrm{F}$. As the bias voltage increases bands with 
$\Delta_5$, $\Delta_{2}$ and $\Delta_{2^\prime}$ symmetry became available for transport. However, these are filtered by 
symmetry by the BiF$_3$ barrier and the transmission remains generally small. $T(E;V)$ for the anti-parallel configuration is
essentially a convolution of those for the majority and minority spins in the parallel one, i.e. it traces closely the minority spin
transmission.  

 \section{Conclusion}
In conclusion, we have identified from all known and predicted Heusler alloys a materials combination, which can act as an alternative 
to the FeCoB/MgO/FeCoB heterostructure. In particular we have looked at the Fe$_3$Al/BiFe$_3$/Fe$_3$Al stack and demonstrated
that this junction operates with the same symmetry spin-filtering mechanism of FeCoB/MgO, and as such can display extremely
high TMR values. Interestingly, the extended $\Delta_1$ spin-gap along the (100) direction of Fe$_3$Al, gives us a large energy window
where to expect a significant TMR. As such for this proposed junction we expect a strong TMR retention at high voltage.

\begin{acknowledgments} 
%
This work was supported by the Higher Education Research Promotion and National Research University Project of Thailand, 
the Office of the Higher Education Commission, the National  Center (NANOTEC), NSTDA, Ministry of Science and Technology, 
Thailand, through its program of Centers of Excellence Network and by Science Foundation of Ireland (14/IA/2624), and financial 
support from the European Union's Horizon 2020 PETMEM and TRANSPIRE projects. 
%
The authors wish to acknowledge the SFI/HEA Irish Centre for High-End Computing (ICHEC) and the Trinity Centre for 
High Performance Computing (TCHPC) for the provision of computational facilities and supports.
\end{acknowledgments}



\end{document}